\documentclass[aps,pre,twocolumn,floatfix]{revtex4}
\usepackage{graphicx}

\begin{document} 
\title{Effect of the Earth's Coriolis force on the large-scale circulation of turbulent Rayleigh-B{\'e}nard convection}
\author{Eric Brown}
\author{Guenter Ahlers}
\affiliation{Department of Physics and iQCD, University of California, Santa Barbara, CA 93106}

\date{\today}
 
Under consideration for publication in Phys. of Fluids
 
\begin{abstract}

We present measurements of the large-scale circulation (LSC) of turbulent Rayleigh-B{\'e}nard convection in water-filled cylindrical samples of heights equal to their diameters.  The orientation of the LSC had an irregular time dependence, but revealed  a net azimuthal rotation with an average period of about 3 days for Rayleigh numbers $R \stackrel{>}{_\sim} 10^{10}$.   On average there was also a tendency for the LSC to be aligned with upflow to the west and downflow to the east, even after physically rotating the apparatus in the laboratory through various angles.   Both of these phenomena could be explained as a result of the coupling of the Earth's Coriolis force to the LSC.  The rate of azimuthal rotation  could be calculated from a model of diffusive LSC orientation meandering with a potential barrier due to the Coriolis force.  The model and the data revealed an additional contribution to the potential barrier that could be attributed to the cooling system of the sample top which dominated the preferred orientation of the LSC at high $R$. The tendency for the LSC to be in a preferred orientation due to the Coriolis force could be cancelled by a slight tilt of the apparatus relative to gravity, although this tilt affected other aspects of the LSC that the Coriolis force did not. 

\end{abstract}
\maketitle

\section{Introduction}

The problem of Rayleigh-B{\'e}nard convection (RBC) consists of a fluid sample heated from below \cite{Si94, Ka01, AGL02}.  In our case the sample was a cylindrical container filled with water.  This system is defined by three parameters:  the Rayleigh number $R \equiv \alpha g \Delta T L^3/\kappa \nu$ ($\alpha$ is the isobaric thermal expansion coefficient, $g$ the acceleration of gravity, $\Delta T$ the applied temperature difference, $L$ the height of the sample, $\kappa$ the thermal diffusivity, and $\nu$ the kinematic viscosity), the Prandtl number $\sigma \equiv \nu / \kappa$, and the aspect ratio $\Gamma \equiv D/L$ ($D$ is the diameter of the sample).  Convection of heat occurs primarily as a result of the emission of volumes of hot fluid known as ``plumes" from  a bottom thermal boundary layer that rise due to a buoyant force, while cold plumes emitted from a top boundary layer sink.  In the turbulent regime of $\Gamma=1$ samples that we studied, these plumes drive a large-scale circulation (LSC) \cite{KH81, SWL89, CGHKLTWZZ89, CCL96, QT01a, FA04, SXT05,TMMS05} which is oriented nearly vertically with up-flow and down-flow on opposite sides of the sample.  

While the LSC configuration does not have the cylindrical symmetry of the sample, we found that the structure underwent an erratic azimuthal motion that allowed it to sample all angles over time. We showed elsewhere that, on a long time scale (days), this azimuthal motion could be partly described as a diffusive process. We believe it to be driven by the intense turbulent fluctuations that prevail throughout the system. Interestingly, relatively sudden re-orientations through various angles (with durations of order minutes)  were superimposed onto the diffusive motion at intervals much longer than their duration.  These re-orientations could occur via two distinct mechanisms. One of these (a ``rotation") consisted of a relatively fast rotation of the circulation plane (or of the vorticity vector of the LSC) without significant change of the circulation speed (or of the magnitude of the vorticity vector). The other (a ``cessation") involved the complete decay of the circulation, followed by a spontaneous new start with a different orientation. These processes and their statistical properties have been discussed in detail in previous work \cite{BA06, BNA05a}.

Over long time scales (days) we found that the LSC did not sample all angles equally, suggesting that the azimuthal diffusion was subject to some asymmetric potential. In addition, the meanderings of the LSC orientation occasionally added up to rotations through approximately $2\pi$, which we refer to as ``revolutions". For $R \agt 10^{10}$ clockwise (when viewed from above) revolutions were found to be more frequent than counter-clockwise ones.  In this paper we report on these measurements, and develop a simple model for the effect of the Coriolis force on the LSC that explains many of these observations. Experimentally, the strongest qualitative evidence for the influence of the Coriolis force on the LSC is a change in the sample frame of the azimuthal location of the preferred orientation that occurred when the entire apparatus was rotated  through a constant angle $\gamma$ in the laboratory frame. 

In order to develop the model, we show that the Earth's rotation applies a torque to the horizontal flow components near the top and bottom plate that by itself and in the northern hemisphere would induce a rotation in the clockwise direction. In the absence of dissipation the torque in turn would be diminished by this rotation, leading to a rotation frequency precisely equal to that required for the  torque to vanish and for the orientation of the LSC plane to be steady in an inertial frame. In the physical system one also needs to consider the dissipation experienced near the side wall and the top and bottom plates; we introduced this dissipation into the model by considering the viscous drag across a laminar boundary layer and found that it reduces the rotation frequency. Next we need to introduce the coupling of the Earth's rotation to the vertical components of the flow near the side wall. It imposes a preferred orientation that is just north of west and that provides an azimuthally varying potential. This potential interferes with  the rotation, and the two effects mentioned so far  would impose a unique LSC orientation near the minimum of the potential. However, the turbulent background fluctuations that we have neglected up to now couple to the LSC and cause a meandering in the neighborhood of the potential minimum, and induce occasional transitions across the potential barrier that result each time in a rotation through approximately one revolution in either direction. The clockwise and counter-clockwise transitions are not equivalent because the coupling to the horizontal velocity components provides a slanting background to the potential (a ``washboard" potential), leading to a preference for rotations in the clockwise direction when viewed from above,  as seen in the experiment. Earlier \cite{BA06} we showed that the LSC meandering has some qualities of a  diffusive process, and derived its azimuthal diffusivity  from measurements of the long-time angular fluctuations of the LSC orientations. We now use these diffusivities, which vary with $R$, in a stochastic model based on the potential discussed above to derive expressions for the probability distribution of the LSC orientation, for total number of rotations, and for the excess number of rotations in the clockwise direction  from a Fokker-Planck equation. These results reproduce many of our experimental measurements.  However, the model does not explain why the net rotation is absent for  $R\stackrel{<}{_\sim} 10^{10}$. At large $R$ we also found that, in addition to the Coriolis force, we had to invoke an additional azimuthal asymmetry (associated with small imperfections of the apparatus) in order to explain all of our data over wide parameter ranges. 

It is interesting to compare our results with experiments by Hart {\it et al.} \cite{HKO02} where turbulent RBC samples of size comparable to ours  were rotated deliberately at a constant rate in the laboratory frame about a vertical axis. That work involved rotation rates $\Omega$ much larger than the Earth's rotation, and therefore imposed much larger Coriolis forces upon the LSC. It showed that the LSC underwent an azimuthal precession in the frame of the rotating sample and in the retrograde direction at a rate which was proportional to $\Omega$ when $\Omega$ was not too large. We also find a net rotation of the LSC in the clockwise (i.e. retrograde) direction. However, when scaled by the axial component of $\Omega$ its rate is larger by a factor of three or so. In this comparison it should be considered that the Earth's rotation rate of $7.3\times 10^{-5}$ rad/s is more than two orders of magnitude smaller than the smallest experimental $\Omega$ values used in Ref.~\cite{HKO02}. A possible explanation for the quantitative difference may be found in a greater role of viscous dissipation at the larger $\Omega$ values used in Ref.~\cite{HKO02}. 

Hart et al. also developed a model for the coupling of the Coriolis force created by their deliberate rotation to their LSC that in some ways is similar to ours. However, in their case the coupling is only to the horizontal velocity component because the rotation axis is parallel to the vertical flow. Their model thus has no potential extrema, the rotation of the LSC is only in the retrograde direction, and stochastic effects need not be invoked to  yield rotation. When we set the latitude equal to $\pi/2$ in our model and neglect stochastic effects, our model becomes equivalent to that of Hart et al., except that those authors used a different approach toward modeling the dissipation.

While the Coriolis force plays a major part in many geophysical and astrophysical convection systems, it is rare that it can be observed in laboratory fluid-flow experiments because of the small magnitude of the force on laboratory scales. One example is high-precision measurements of the velocity profile of laminar flow in a long pipe \cite{DN98}.  It was found that the expected parabolic profile of the velocity field was skewed by the Earth's rotation. In that case it was possible to carry out a quantitative calculation of this effect based on the equation of motion (the Navier-Stokes equation) of the system because the unperturbed velocity was known analytically, the velocity field remained time independent, and the Coriolis-force perturbation could be calculated at linear order.
For our system the turbulent flow field is much more complicated and a full treatment based on the equation of motion does not seem feasible except perhaps by direct numerical simulation. Thus we resorted to the simplified model described above.

A careful study of the effects of the Earth's Coriolis force could prove useful for understanding and analyzing data in other high-precision RBC experiments.   For instance, our study made it possible to explore other asymmetries that affected the LSC in our experiment, in particular effects due to our top-plate cooling-system and to small deliberate misalignments relative to gravity \cite{ABN05}.

In the next section we describe briefly the experimental measurements; a more detailed account of the apparatus had been provided before \cite{BNFA05}. Section \ref{sec:rotation_expt} gives the results accumulated over many months for the azimuthal rotation of the LSC. Data for the diffusivity of the LSC meandering are given in Sect.~\ref{sec:diff}. The measurements of the probability distribution of the LSC orientation are discussed in Sect.~\ref{sec:orientation_expt}. Measurements of the instantaneous rotation rate turn out to be useful to determine a parameter of the model, and thus are given in Sect.~\ref{sec:inst_rot_rate}. In Sect. \ref{sec:model} we present a model for the interaction between the Earth's Coriolis force, the LSC, and the background turbulence of the system. In a number of subsections we derive predictions for several properties of the LSC and compare them with experimental results. The properties include the preferred orientation, the effect of dissipation, the probability distribution of the LSC orientation, the net rotation rate, and the total rate of occurrence of revolutions. Comparison with other experiments is provided as well. 
The influence of azimuthal imperfections in the sample cell,  we believe primarily due to the cooling system of the top plate, is incorporated in the model in Sect.~\ref{sec:asymmetry}. Section \ref{sec:tilt} considers the influence of a slight misalignment relative to gravity of the sample, and discusses how this tilt can in part cancel the Coriolis-force effects. A summary and some conclusions are presented in Sect.~\ref{sec:summary}.

\section{Experimental Method}
\label{sec:experiment}

The experiments were done with two cylindrical samples with aspect ratio $\Gamma  \approx 1$ that were the medium and large sample described in detail elsewhere \cite{BNFA05}.  Both had circular copper top and bottom plates with a plexiglas side wall that fit into a groove in each plate.  The plate surfaces that touched the fluid sample were nominally symmetric and fine machined to a roughness of less that 2 micro-m. The side walls had variations of their diameter of less than $10^{-2}$ cm.  There was a small semi-circular hole 1.6 mm in diameter at the edge of each plate that was used for filling and to allow expansion of the fluid.  There were no internal sensors or other structures that could lock the flow direction.  A resistive heater wire was embedded in the bottom plate in such a way that it covered all regions of the plate equally for even heating.  The top plate contained a double-spiral water-cooled channel which covered all regions of the plate, with inlet and outlet on opposite sides of the plate. The medium sample had dimensions $D=24.81$ cm and $L=24.76$ cm.  The large sample had $D = 49.67$ cm and two side walls which could be interchanged with $L=50.61$ and $L= 49.54$ cm.  Each apparatus was filled with de-ionized de-gassed  water and the average temperature of the bottom and top plates was kept at $40.0^{\circ}$ C where $\sigma = 4.38$. The two samples of different physical size allowed us to cover a larger range of $R$ at the same $\sigma$ and $\Gamma$, so the overall range studied was $3\times10^8 \stackrel{<}{_\sim} R \stackrel {<}{_\sim} 10^{11}$.    Both samples were carefully levelled to better than 0.001 rad, except for the experiments in which we deliberately tilted the samples.

Eight blind holes, equally spaced azimuthally in the horizontal mid-plane of the samples,  were drilled from the outside into each side wall. Thermistors were placed into them so as to be within $d = 0.07$ cm of the fluid surface. They were numbered $i = 0, \ldots,7$ in the counter-clockwise direction when viewed from above, with thermistor 0 normally located in an easterly position except at times when the entire apparatus was rotated in the clockwise direction through an additional angle $\gamma$.  Since the LSC carried warm (cold) fluid from the bottom (top) plate up (down) the side wall, these thermistors detected the location of the upflow (downflow)  of the LSC by indicating a relatively high (low) temperature.  No parts of the thermistors extended into the sample where they might have  perturbed the flow structure of the fluid.

We made measurements of the thermistor temperatures at the horizontal midplane at time intervals $\delta t \approx 6$ seconds, and fit the function 
\begin{equation}
T_i = T_0 + \delta\cos(i \pi/4 - \theta_0), i = 0, \ldots,7\ ,
\end{equation}
 separately at each time step, to these eight readings. Examples of such fits have been shown previously \cite{BA06, BNA05a}.  The fit parameter $\delta$ is a measure of the amplitude of the LSC and $\theta_0$ is the azimuthal orientation of the plane of the LSC circulation in the sample frame.   As defined here, the orientation $\theta_0$ is on the side of the sample where the LSC is up-flowing and is measured relative to the location of thermometer zero.  Typically the uncertainties for a single measurement were about 13\% for $\delta$ and 0.02 rev. for $\theta_0$.   When we fit the side-wall thermistor-temperatures to get $\theta_0$, the fit only determined the value of $\theta_0$ modulo $2\pi$. In order to track $\theta_0$ continuously through many azimuthal rotations we chose the value that was within $\pi$ of the value measured at the previous time step.  

\section{Experimental Results}
\label{sec:results}

\subsection{Revolutions and net rotation of the LSC}
\label{sec:rotation_expt}

\begin{figure}
\includegraphics[width=3in]{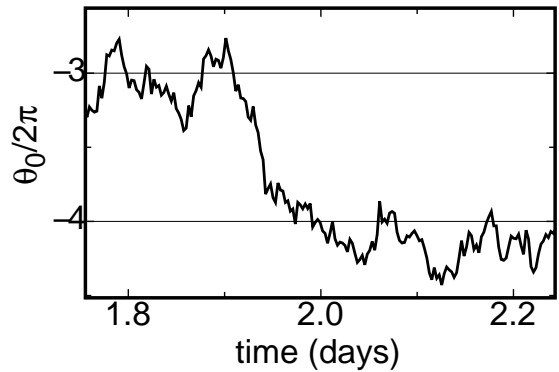}
\caption{An illustration of a revolution. The figure shows a time series of the orientation $\theta_0(t)$ for the large sample over a time interval of about 12  hours (1/2 day) at $R=8.9\times10^{10}$.}
\label{fig:rev}
\end{figure}

Elsewhere we presented measurements of sudden re-orientations of the plane of circulation of the LSC through a continuum of possible angles $\Delta \theta$ \cite{BA06}. They were relatively rare events in that their collective duration took up only a small fraction of the observation time. They assumed the form of either cessations, where the flow actually stopped before re-starting in a new randomly chosen orientation, or of rotations where the flow continued but its vorticity vector rotated relatively rapidly through some angle $\Delta \theta$. Rotations and Cessations typically occurred rather quickly, in the time span of about one turnover time of the LSC (which is on the order of $100$ s).  Here we focus on yet another event which was like a rotation in the sense that the LSC orientation changed without a significant change of the LSC speed, but the change in orientation $\Delta \theta$ was close to $2\pi$ and required a larger time interval, typically of an hour or so (of order $30$ turnover times).  
 These events will be referred to as ``revolutions". An illustration of a revolution is given in fig.~\ref{fig:rev}. After a revolution  the orientation (modulo $2\pi$) is essentially the same as it was before that event. In our previous analysis \cite{BA06} revolutions were too slow to be counted as rotations.
 
We shall show below that revolutions are associated with diffusive motion in a periodic potential with  a minimum at a particular azimuthal orientation. This potential is  created to a large extent by the coupling of the Earth's Coriolis force to the LSC. It turns out that clockwise revolutions are more frequent than counter-clockwise ones, leading to a net rotation over long time periods. This asymmetry again can be explained in terms of the Coriolis force.  

\begin{figure}
\includegraphics[width=3in]{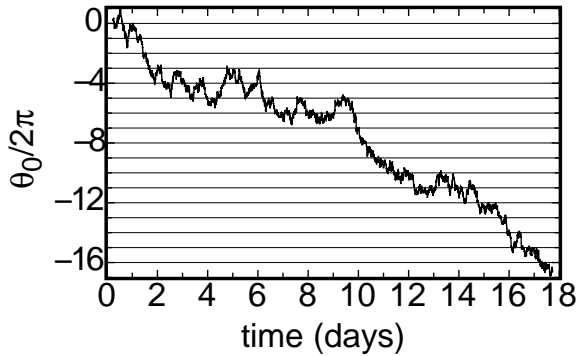}
\caption{A time series of the orientation $\theta_0(t)$ for the large sample over 18 days at $R=8.9\times10^{10}$.  There was a net rotation of 17 clockwise revolutions. }
\label{fig:net_rotation}
\end{figure}

An example of net rotation of the LSC for the large sample at $R=8.9\times10^{10}$ is seen in a time series of $\theta_0(t)$ shown in Fig.~\ref{fig:net_rotation}.  There is a net rotation of 17 clockwise revolutions over 18 days.  Inspection of the time series shows that the LSC had a tendency to remain aligned near a preferred orientation $\theta_m$ much of the time,  but that it occasionally underwent significant positive or negative rotations before locking in again near $\theta_m$ (modulo $2\pi$) as illustrated in Fig.~\ref{fig:rev}.

To facilitate automated data analysis, we define a positive revolution as an event where  $\theta_0(t)$ rotates from the range $\theta_m -\pi/4 < \theta_0 < \theta_m + \pi/4$ to the range $(\theta_m + 2\pi) - \pi/4 < \theta_0 < (\theta_m + 2\pi) + \pi/4$.    A negative revolution is the analogous change of approximately $-2\pi$.  We require the orientation to  come within a small range of $\theta_m$ to make sure that each revolution counted is a rotation of approximately $2\pi$ and not jitter around an angle $\theta_m\pm \pi$. Let us describe the sequence of $N$ revolutions by a sequence of $N$ numbers $r_i$ which are equal to 1 (-1) for a counter-clockwise (clockwise) revolution. The net rotation $\cal R$ is then equal to the sum of the $N$ values of $r_i$. For every adjacent pair of values $r_i$ and $r_{i+1}$ we compute the product $d_i = r_i r_{i+1}$. The parameter $d_i$ will be equal to 1 if the two members of the pair of $r_i$ correspond to rotations in the same direction, and equal to -1 otherwise. The sum $\cal D$ of all ${N} - 1$ $d_i$ is a measure of the correlation between successive events.
   
Counting the number of revolutions over all of the experimental running time in the large sample of 258 days, covering  the range $3\times 10^9 \stackrel{<}{_\sim} R  \stackrel{<}{_\sim} 10^{11}$,   we found a net rotation ${\cal R} = -77$, corresponding to 77 clockwise revolutions out of a total of ${N} = 439$ revolutions.  The net correlation between successive events over this time was ${\cal D} = 1$; the smallness of this number relative to the total number of revolutions $N$ shows that the directions of successive revolutions were not correlated with each other, and that the net rotation was not due to a long time scale over which the LSC rotated in one direction.  It also allows us to use binomial statistics to analyze the significance of the results.  If there were a 50\% chance for each revolution to be either in the positive or negative direction, we would expect that the net rotation after ${N} = 439$ revolutions would be less than about one standard deviation $\sigma_{N} \equiv \sqrt{{N}}$ which is approximately 21 revolutions.  Thus  the observations of 77 clockwise revolutions, corresponding to 3.7 $\sigma_{N}$ away from a mean of zero, indicate that the net rotation is a real effect with a confidence limit of 99.98\%.  The average rate of net rotation in the large sample was $\omega_{\cal R} = -0.30 \pm 0.08$ rev. per day (clockwise is negative in our coordinates).  In the medium sample,  covering the  range $3\times 10^8 \stackrel{<}{_\sim} R  \stackrel{<}{_\sim} 10^{10}$, there were only 9 net counter-clockwise revolutions out of a total of 331 revolutions over 194 days, and ${\cal D} = -8$ for an average rate of $\omega_{\cal R} = 0.05 \pm 0.09$ rev. per day.  One sees that the results from the medium sample are consistent with no net rotation, and at the least they indicate a smaller rotation rate than was found in the large sample. 

We compare our data to an experiment by Hart {\it et al.} \cite{HKO02} with $\Gamma = 1$ and Prandtl number $\sigma = 8.4$ that studied the effect on the LSC of the Coriolis force due to rotating a convection apparatus at various applied rotation rates $\Omega$ over the range $0.015$ rad/s $  \stackrel{<}{_\sim}\Omega  \stackrel{<}{_\sim} 0.7 $ rad/s. Two samples were used; one was about the same size as our large and another a little smaller than our medium sample.  For their large sample the authors reported no net rotation of the LSC for $R \stackrel{<}{_\sim} 10^{10}$ but a net rotation of $\omega_{\cal R} = -0.23\Omega$ in the rotating sample frame for $R \simeq 2.9 \cdot 10^{11}$ \cite{HKO}.  This is seemingly consistent with our rotation rate of $\omega_{\cal R} = (-0.30 \pm 0.08)\Omega$, even though their applied rotation rates were several orders of magnitude larger than our ({\it i.e.} the Earth's) rotation rate $\Omega = 7.3\times 10^{-5}$ rad/s, although later we will show that such a straightforward comparison is probably not correct.  It is also interesting that neither experiment found net rotation in the smaller sized apparatus, which yielded data only for the range $R \stackrel{<}{_\sim} 10^{10}$.

\begin{figure}
\includegraphics[width=3in]{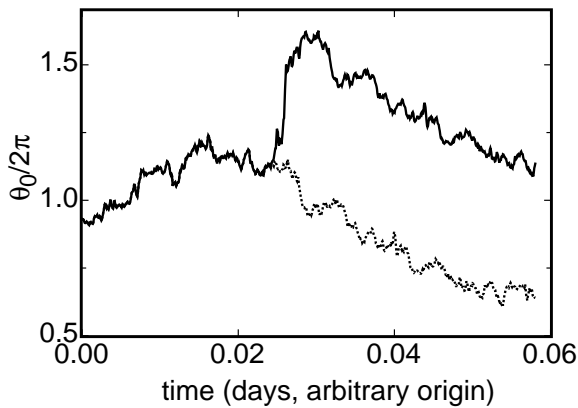}
\caption{Time series of the orientation $\theta_0(t)$ at $R=4.7\times10^{10}$ for the large sample. The solid line is the original data. The dotted line is the result of removing  a re-orientation by shifting the data beyond it along the horizontal and vertical axes.}
\label{fig:filter}
\end{figure}

\subsection{Azimuthal diffusion of the LSC}
\label{sec:diff}

The re-orientations (cessations and rotations) were relatively rare and their duration occupied a relatively short fraction of a given long time series. During the remainder of the time the azimuthal orientation $\theta_0$ meandered in a seemingly random manner.  Elsewhere we reported that over long time scales  the orientation meanderings have the diffusive quality that the root-mean-square (rms) rotation rate computed over various time intervals $n\delta t$ is given by 
\begin{equation}
\dot\theta^{rms}_n = \sqrt{\frac{D_{\theta}}{n\delta t}}\ .
\label{eqn:diff}
\end{equation}
Here $\dot\theta^{rms}_n = \sqrt{\langle[\theta_0(t+n\delta t) - \theta_0(t)]^2\rangle}/(n\delta t)$, $D_{\theta}$ is a diffusivity, $\delta t$ is the time interval between successive data points, $n$ is a positive integer, and $\langle\rangle$ indicates a time average  \cite{BA06}.
We found an effective diffusivity given by  $D_{\theta} = 4.45\times10^{-4} R^{0.556}\times \nu/L^2$ \cite{BA06}. Since re-orientations do not occur very frequently, that analysis had used time series of $\theta_0$ that included all re-orientations even though these events could not be described by simple diffusion. Motivated by the present need for more accurate values of $D_{\theta}$, we repeated the analysis after filtering our re-orientations. The procedure is illustrated in Fig.~\ref{fig:filter}. There the solid line represents a short section of the original time series. It clearly reveals a re-orientation. The modified time series is shown as the dotted line. The displacements along both the time and the $\theta_0$ axis during the re-orientation have been removed by shifts parallel to the axes. 

\begin{figure}
\includegraphics[width=3in]{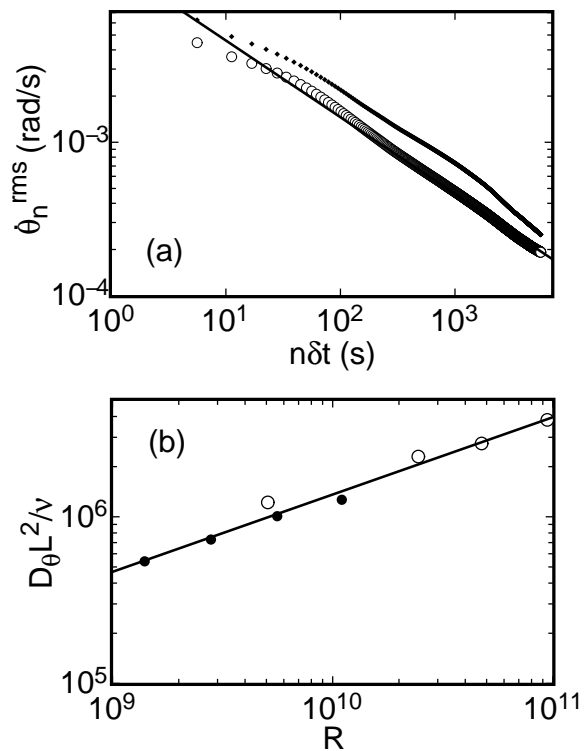}
\caption{(a): The rms rotation rate $\dot\theta^{rms}_n$ as a function of $n\delta t$ on logarithmic scales for the large sample and for $R = 5\times 10^9$. Solid symbols: from the analysis of the complete data sequence. Open symbols: from the analysis of the data sequence after removal of re-orientations. The solid line is a fit of Eq.~\ref{eqn:diff} to the open symbols, adjusting $D_\theta$. (b): The diffusivities $D_\theta$, scaled by $L^2/\nu$, as a function of $R$ for the medium (solid circles) and the large (open circles) sample as determined after removal of re-orientations.}
\label{fig:D}
\end{figure}

Figure~\ref{fig:D}a shows an example of $\dot\theta^{rms}_n$ as a function of $n\delta t$ on logarithmic scales for  both methods of analysis. One sees that omitting the re-orientations still yields data consistent with Eq.~\ref{eqn:diff} (solid line in the figure), but that the diffusivity $D_\theta$ is about a factor of two smaller when the re-orientations are omitted.
The new analysis yielded the results for $D_\theta$ shown in Fig.~\ref{fig:D}b. They can be represented well by
\begin{equation}
D_{\theta} = 2.3\times10^{-3} R^{0.46 \pm 0.03}\times \nu/L^2\ .
\label{eqn:D_theta}  
\end{equation}
as is shown by the solid line. It is this diffusivity based on the modified time series for $\theta_0$ that we use in the calculations presented in this paper. Presumably it better represents diffusive processes of $\theta_0(t)$, as the values reported earlier were "contaminated" by non-diffusive reorientations.

\subsection{Preferred orientation of the LSC}
\label{sec:orientation_expt}

\begin{figure}
\includegraphics[width=3in]{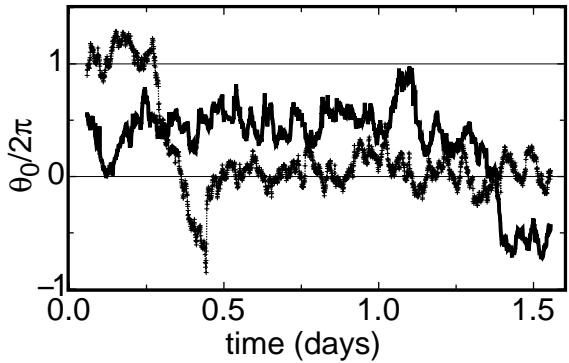}
\caption{Time series of the orientation $\theta_0(t)$ for the large sample at $R=9.4\times10^9$ for sample orientations $\gamma = 0$ (thick solid line) and $\gamma = 0.97\pi$ (thin dotted line).  While there were many fluctuations of $\theta_0(t)$, notice that for each run there was a preferred orientation that the LSC tended to stay near much of the time, and that this orientation shifted by about $\pi$ in the sample frame when the sample was rotated by $\pi$ , indicating that the preferred orientation was due to an external field.}
\label{fig:theta_t}
\end{figure}

  Over the course of this work, the entire apparatus was rotated through several different angles $\gamma$. We measured $\gamma$ as a clockwise angle when viewed from above in the laboratory frame, with $\gamma=0$ when $\theta=0$ was pointing east.  After each such reorientation the top plate of the sample was carefully levelled to within 0.001 rad.  A pair of time series $\theta_0(t)$ before and after one such rotation is shown in Fig.~\ref{fig:theta_t} for $R=9\times 10^{9}$ in the large sample.  The solid line corresponds to $\gamma = 0$ and the dotted one is for $\gamma = 0.97\pi$.  While there were many fluctuations in $\theta_0(t)$ for both time series, one sees that for each run there was a preferred orientation, and that this orientation shifted by about $\pi$ in the sample frame when the sample was rotated by $\gamma \simeq \pi$, suggesting that this preferred orientation was due to an external field fixed in the laboratory frame and not due to an asymmetry fixed in the sample frame.

\begin{figure}
\includegraphics[width=3in]{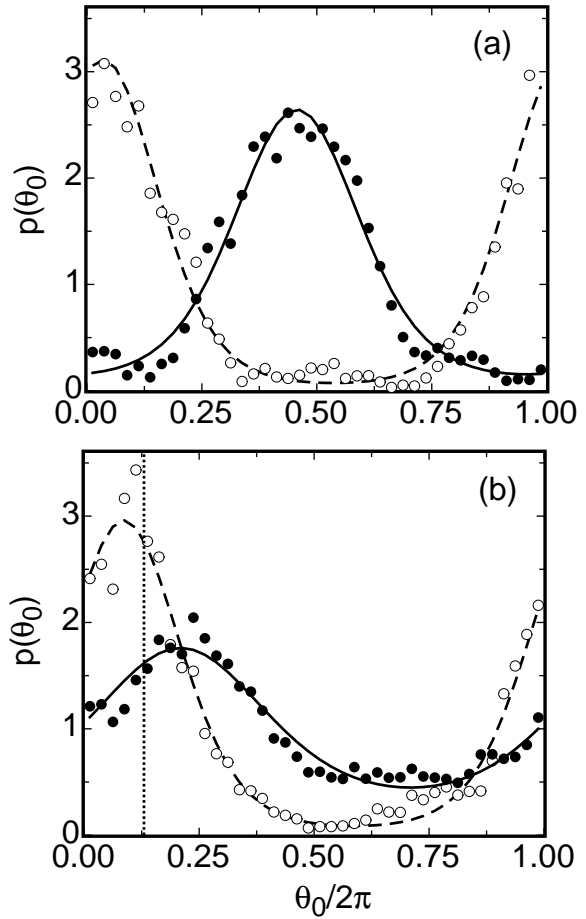}
\caption{(a):  The probability distribution of the orientation $p(\theta_0)$ for the data shown in Fig.~\ref{fig:theta_t}.  The function $p(\theta_0) = C + (1-C)/(\sqrt{2\pi}\sigma_{\theta})\times\exp[-(\theta_0-\theta_m)^2/(2\sigma_{\theta}^2)]$ was fit to the data to get the preferred orientation $\theta_m$. Solid circles: $\gamma = 0$, fitted by solid line.  Open circles: $\gamma = 0.97\pi$, fitted by dashed line. (b): $p(\theta_0)$ as in (a), but for $R= 9.4\times10^{10}$.   The vertical dotted line corresponds to the ``hot" side of the top plate (see section Sect.~\ref{sec:asymmetry} for an explanation). }
\label{fig:theta_pref_fit}
\end{figure}

We determined the preferred orientations $\theta_m$ of the LSC from the probability distributions of the orientations $p(\theta_0)$.  The Gaussian function

\begin{equation}
p(\theta_0) = \frac{C}{2\pi} + \frac{1-C}{\sqrt{2\pi}\sigma_{\theta}}\exp\frac{-(\theta_0-\theta_m)^2}{2\sigma_{\theta}^2}
\label{eqn:ptheta_gauss}
\end{equation}

\noindent was fitted to the data for $p(\theta_0)$.  Examples of such fits are shown in Fig.~\ref{fig:theta_pref_fit}a for the data of Fig.~\ref{fig:theta_t} and in Fig.~\ref{fig:theta_pref_fit}b for data with the same $\gamma=0$ and $\gamma=0.97\pi$ but with $R = 9.4\times 10^{10}$.  Although the coordinate $\theta$ is periodic and the fitting function is not, this did not cause any difficulties since we found that the width of the Gaussian generally was much less than $\pi$.  The only correction needed was a shift of the values of $\theta$ by multiplies of $2\pi$ so that $\theta_0$ was within $\pm \pi$  of $\theta_m$.

In Fig.~\ref{fig:theta_pref_fit}a the shift of $\theta_m$ [the peak of $p(\theta_0)$] after rotating the apparatus is unmistakable. One can see that it is only a little smaller than $\gamma$. However, at larger $R$, the shift of $p(\theta_0)$ was much smaller than $\gamma$. This is shown in Fig.~\ref{fig:theta_pref_fit}b for $R=9.4\times 10^{10}$, for the same $\gamma=0.97\pi$.  The smaller shift suggests that the preferred orientation at this $R$ is chosen in part by some sample asymmetry rather than exclusively by an external field. This issue will be discussed in detail in Sect.~\ref{sec:asymmetry}.      

\begin{figure}
\includegraphics[width=3in]{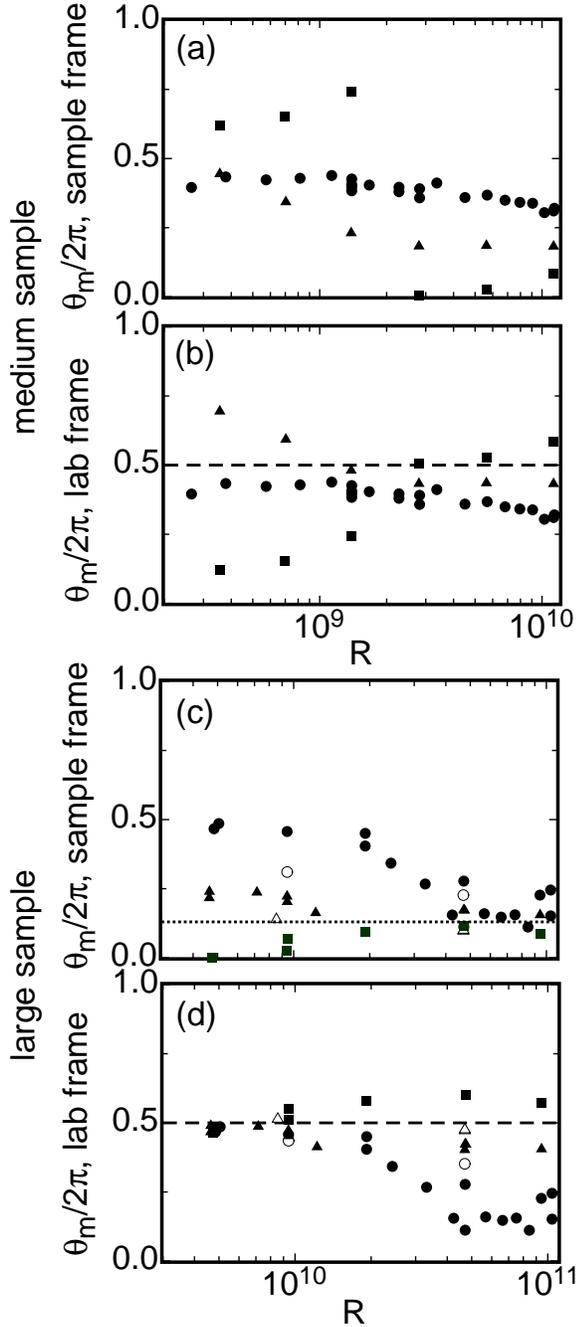}
\caption{The preferred orientation $\theta_m$ of the LSC vs. $R$ for various rotations  in the laboratory frame of the samples through angles $\gamma$. (a): medium sample, sample reference frame. (b): medium sample, laboratory reference frame.  (c): large sample, sample reference  frame.  (d): large sample, laboratory reference frame.  $\theta_m = 0$ is east in the laboratory reference frame when $\gamma = 0$. Solid circles:  $\gamma = 0$.  Open circles:  $\gamma = 0.25\pi$.  Solid triangles:  $\gamma = 0.50\pi$.  Open triangles:  $\gamma = 0.75\pi$. Solid squares:  $\gamma = 0.97\pi$. Dashed line: west.  Dotted line:  the ``hot" side of the top plate (see Sect.~\ref{sec:asymmetry} below).}
\label{fig:theta_pref}
\end{figure}

The results for $\theta_m$ are shown in Fig.~\ref{fig:theta_pref}, in both the sample reference frame (a and c) and the laboratory reference frame (b and d).  One sees clearly for the large sample in the sample frame (Fig.~\ref{fig:theta_pref}c) that the data for different angles $\gamma$ approach each other at large $R$, indicating that there is an asymmetry associated with the apparatus that dominates $\theta_m$ at large $R$. Comparing with the vertical dotted line in Fig.~\ref{fig:theta_pref_fit} one sees that the preferred angle in the sample frame at large $R$ is close to the hot side of the top plate. This will be discussed in detail in Sect.~\ref{sec:asymmetry}.   

For the large sample in the laboratory frame (Fig.~\ref{fig:theta_pref}d), we see that the data for different $\gamma$ nearly collapse at small $R$, indicating that there is an asymmetry associated with an external field, fixed in the laboratory frame,  that causes $\theta_m$ to align to the west.  Similar trends can be seen in the medium sample, but the data do not collapse well, suggesting that the asymmetries are weaker or more equally balanced there.  

\begin{figure}
\includegraphics[width=3in]{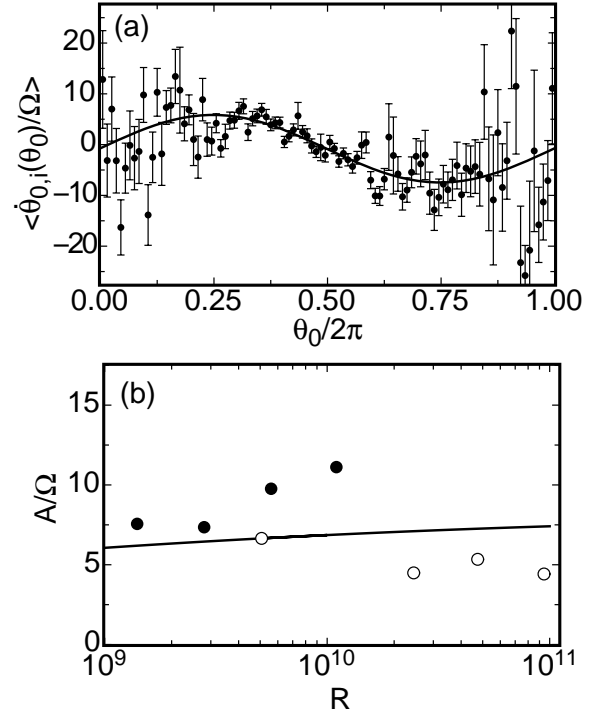}
\caption{(a): The average of the instantaneous azimuthal rotation rate $\langle\dot\theta_{0,i}(\theta_0)\rangle$ as a function of $\theta_0$ at $R=5\times10^9$ in the large sample.  Solid line:  a fit of $\langle\dot\theta_{0,i}(\theta_0)\rangle = A\sin\theta_0 - B$ to the data.  (b): The fit amplitude $A$ vs. $R$.  Solid circles:  medium sample.  Open circles: large sample. Solid line:  model value  $A = 2a\Omega\cos\phi/(12R_e^{-1/2} +1)$ with $a = 5.0$.}
\label{fig:dtheta_theta}
\end{figure}

\subsection{The instantaneous rotation rate of the LSC}
\label{sec:inst_rot_rate}

The model for the influence of the Coriolis force on the LSC to be presented in Sect.~\ref{sec:model} involves one free parameter. Within the context of that model this parameter can be determined from independent measurements of the time averaged  {\it instantaneous} rotation rate $\langle\dot \theta_{0,i}(\theta_0)\rangle \equiv \langle[\theta_{0}(t+\delta t) - \theta_{0}(t)]/\delta t\rangle$ where $\langle\rangle$ denotes an average over time but is left as a function of $\theta_0$.  The purpose of the time average is to remove stochastic effects so what remains represents a deterministic rotation rate. In Fig.~\ref{fig:dtheta_theta}a  we present experimental results for $\langle\dot \theta_{0,i}(\theta_0)\rangle$. At each $\theta_0$ it shows an average over many data points of $\dot\theta_{0,i}(\theta_{0})$ for $R=5\times10^{9}$ in the large sample, where $\theta_m$ was found to be fixed in the laboratory frame (see Fig.~\ref{fig:theta_pref}d) and presumably determined primarily by the Coriolis force (as discussed in Ref. \cite{BA06}, $\dot\theta_{0,i}$ was calculated with a slight time delay $\tau$ after $\theta_0$ was reached which satisfied the condition $\ddot\theta_0 = 0$).  Below in Sect.~\ref{sec:dissipation} we show that our model predicts the relationship  $\langle \dot\theta_{0,i}(\theta_{0})\rangle =A\sin\theta_0 + B$. This function was fit to the data for $\langle\dot\theta_{0,i}(\theta_0)\rangle$ by adjusting $A$ [$B$ is a known constant, see Eq.~\ref{eqn:ab}].   The fit amplitude $A$ vs. $R$ is shown in Fig.~\ref{fig:dtheta_theta}b, along with a solid line representing a model value for $A$ derived from Eq.~\ref{eqn:ab} and to be discussed below.  For larger $R$ where the Coriolis force does not determine the preferred orientation, we inserted a second free parameter into the fitting function to allow for a shift in $\theta_m$.  This issue will be discussed in Sect.~\ref{sec:asymmetry}.

\section{A model for the influence of the Earth's Coriolis force on the LSC}
\label{sec:model}

\subsection{Preferred orientation}
\label{sec:orientation}

The Earth's Coriolis force can be shown to be responsible for the net rotation of the LSC, and to dominate the preferred orientation for some parameter ranges.  The contribution of the Coriolis force to the fluid acceleration in the sample frame is given by
\begin{equation} 
\frac{d\vec u}{dt} = -2\left(\vec\Omega + \frac{d\vec\theta_0}{dt}\right)\times \vec u - \frac{d^2\vec\theta_0}{dt^2} \times \vec r\ .
\label{eq:dudt}
\end{equation}
Here $\vec u$ is the fluid velocity, $\Omega$ is the Earth's rotation rate of one revolution per day, and $\dot\theta_0=d\theta_0/dt$ is the azimuthal rotation rate of the LSC in the frame of the sample.  The sum $\vec \Omega + d\vec\theta_0/dt$ is the total rotation rate in an inertial frame.  The centrifugal force is ignored because it is of 2nd order in $\Omega$, which is a small parameter.  We shall take $\phi = 34.46^{\circ}$, the angle of latitude at our location, to be positive in the northern hemisphere.
It is natural to consider $\vec u = (u_r,u_\theta,u_z)$ in the cylindrical coordinates of the sample. 

Although the experimental data for the LSC orientation $\theta_0$ in Figs.~\ref{fig:net_rotation} to \ref{fig:theta_t} clearly show, on three different time scales,  that  this angle varies erratically in time, we shall ignore the stochastic effects for now as we derive the deterministic effects of the Coriolis force.  If we assume that the deterministic forces are independent of the time-dependent stochastic forces, then the deterministic equations are equivalent to time-averaged equations.  Starting in Sect.~\ref{sec:ptheta} below we will consider the influence of stochastic effects, due to the turbulent background fluctuations, on the flow.

For simplicity we model the LSC as having a square vertical cross section and as consisting of four components:  two horizontal for flow across the top and bottom plates and two vertical near the side wall.  The flow is taken to be horizontal and flowing towards $\theta_0$ in a cone based on the bottom plate and with its apex at the center of the cell.  It is in the opposite direction in a congruent cone based at the top plate with its apex also at the cell center.  
The horizontal flow is assumed to have velocity $u_h = 2zU/L$, where $z$ is the axial coordinate measured from the center of the cell, and $U$ is the velocity at the peak of the profile near the thin boundary layer at $z \simeq \pm L/2$.  The flow in the remainder of the sample is taken to be vertical with velocity $u_v =  2rU/L$, and up-flowing at angles within $\pm \pi/2$ of $\theta_0$, and down-flowing at other angles \cite{BA06}.  The velocity scale $U$ is much larger than $r\dot\theta_0$, so we can safely ignore the rotational contribution to the flow velocities here.

The horizontal components of the flow are deflected to the right in the northern hemisphere, so both the top and bottom legs of the LSC are pushed into a clockwise acceleration: 
\begin{equation}
\dot u_{\theta,h}(z)  = -2u_{h}(z)(\Omega\sin\phi + \dot\theta_0) - \ddot\theta_0r\ .
\label{eq:udot_h}
\end{equation}
We ignore forces in directions other than the azimuthal because they will be balanced by the wall, although they may contribute to some deformation of the LSC structure.  For the vertical components, the upflow is deflected to the west while downflow is deflected to the east in the northern hemisphere.  Since we defined the LSC orientation so that there is upflow at $\theta_0$, the LSC will be pushed to orient $\theta_0$ towards the west ($\theta=\pi$)  according to 
\begin{eqnarray}  
  \dot u_{\theta,v}(r) &=& -2u_{v}(r)\Omega\cos\phi\sin(\pi-\theta_0) - \ddot\theta_0r\nonumber \\
  &=& 2u_{v}(r)\Omega\cos\phi\sin\theta_0 - \ddot\theta_0r\ .
  \label{eq:udot_v}
\end{eqnarray}  
 
We can combine Eqs.~\ref{eq:udot_h} and \ref{eq:udot_v}, taking an average of the stated velocity profiles over the entire volume of the sample to obtain the average azimuthal acceleration.  At this point we also introduce an adjustable parameter $a$ to be used as a fitting coefficient, and obtain
\begin{equation}
 \langle\dot u_{\theta}\rangle = U\left[a\Omega\cos\phi\sin\theta_0 - (\Omega/2)\sin\phi -\dot\theta_0/2\right] - \ddot\theta_0\langle r\rangle\ .
 \label{eqn:ucor}
 \end{equation}
  In doing this velocity average we assumed that $u(r,\theta,z)$ is not correlated with $\dot\theta_0$ or $\sin\theta_0$.  The velocity average led to the second and third terms being half as strong as the first because the horizontal flow in the cone regions takes up $1/3$ of the volume of the sample, and thus a volume half as large as that of the vertical flow region. The parameter $a$ accounts for any other differences in the ratio of vertical to horizontal flow forces; if the model is at all reasonable, then a fit of its predictions to the data should yield a value of $a$ that is of order one. By dividing by the spatially averaged radius $\langle r\rangle = L/3$, and moving the other $\ddot\theta_0$ to the left side, we can now write a single differential equation for $\theta_0$:
 \begin{equation}
\ddot \theta_0 \approx \frac{\langle\dot u_\theta\rangle}{\langle r\rangle } = \frac{3U}{4L}\left[2a\Omega\cos\phi\sin\theta_0 - \Omega\sin\phi - \dot\theta_0\right]
\label{eqn:coriolis}
\end{equation}

\noindent We resorted to doing spatial averages because the forcing is not linear in $r$ and thus there will be some distortion of the LSC structure which we do not consider explicitly.  
We derived $\ddot\theta_0$ as a characteristic angular acceleration representing the entire LSC.

  The first term in the bracket on the rhs of Eq.~\ref{eqn:coriolis} orients the preferred orientation $\theta_m$ to the west, while the second term forces the LSC clockwise, moving the preferred orientation to the north of west.  In Fig.~\ref{fig:theta_pref}, the data for $\theta_m$ at various azimuthal sample orientations $\gamma$ nearly collapse in the laboratory frame at small $R$ in the large sample, and $\theta_m$ is found to be slightly north of west; so the model has successfully predicted the quadrant of $\theta_m$.  Equation \ref{eqn:coriolis} has two equilibrium solutions, corresponding to $\ddot \theta_0 = \dot \theta_0 = 0$, that yield $\theta_m$. They satisfy $\sin\theta_m = (1/2a)\tan\phi$, of which the value of $\theta_m$ in the second quadrant (between north and west) satisfies $d\ddot\theta_0/d\theta < 0$ and thus is a stable equilibrium. However, because of systematic uncertainties in the measurement of $\theta_m$ as well as because of possible influences from sample imperfections that have not yet been included in the model, we can not accurately obtain the value of $a$ at this point. In the absence of noise we would expect the LSC to assume a uniquely defined orientation given by $\theta_m$. We attribute the distribution of $\theta_0$ about $\theta_m$ shown by the data in Fig.~\ref{fig:theta_pref_fit} to the action of the intense turbulent background fluctuations on the LSC.

\subsection{Effect of dissipation}
\label{sec:dissipation}

While Eq.~\ref{eqn:coriolis} provides a differential equation for $\theta_0$, it does not take into account viscous drag, which we consider here.  We make the assumption that the dynamics are strongly damped, so we can ignore the inertial term ($\ddot\theta=0$). Then we can predict the steady state azimuthal rotation rate $\dot \theta_0(\theta_0)$ by assuming that the Coriolis force is balanced by viscous drag in a boundary layer of width $\lambda$.    The average deceleration $\langle\dot u_{\theta,\nu}\rangle$ of the LSC can be estimated as the drag in the boundary layer times the fractional volume of the boundary layer in the sample given by  $\lambda/L$ for each plate and $4\lambda/L$ for the side for a total of $6\lambda/L$:

\begin{eqnarray}
\langle\dot u_{\theta,\nu}\rangle = \langle\nu\nabla^2 u_{\theta}\rangle &\approx& \nu\frac{1}{\lambda^2}\frac{ L\dot\theta_0}{2}\frac{6\lambda}{L} \nonumber  \\
&\approx& \frac{6\nu\dot\theta_0 R_e^{1/2}}{L}
 \label{eqn:udrag}
 \end{eqnarray}

\noindent where in Eq.~\ref{eqn:udrag} we expressed $\lambda \simeq 0.5L (R_e)^{-1/2}$ \cite{GL02} in terms of  the Reynolds number $R_e = UL/\nu$. To find the steady-state rotation rate $\dot\theta_0$ characteristic of the LSC, we equate Eq.~\ref{eqn:ucor} and Eq.~\ref{eqn:udrag}  to obtain

\begin{equation}
 \dot\theta_0 =  A\sin\theta_0 - B
 \label{eqn:dthetadt}
\end{equation}

\noindent where

\begin{eqnarray}
A=\frac{2a\Omega\cos\phi}{12R_e^{-1/2} + 1} \mbox{ and }
B = \frac{\Omega\sin\phi}{12R_e^{-1/2} + 1} \ .
\label{eqn:ab}
\end{eqnarray}

\noindent As already seen above from Eq.~\ref{eqn:coriolis}, the preferred orientation $\theta_m$ is found when $\dot\theta_0$ vanishes, given by $\sin\theta_m = (1/2a)\tan\phi = B/A$. 

\subsection{Instantaneous rotation rate}
\label{sec:rot_rate}

In order to determine the free parameter $a$  of the model, we examine data for the
time averaged  {\it instantaneous} rotation rate $\langle\dot \theta_{0,i}(\theta_0)\rangle$ given in Sect.~\ref{sec:inst_rot_rate} and equate it with $\dot \theta_0(\theta_0)$ of the model, which at this point still does not take into account stochastic effects and thus is equivalent to a time-average.   At $R=5\times10^9$ in the large sample, where the Coriolis force was seen to determine the preferred orientation, we fit Eq.~\ref{eqn:dthetadt} to the data (shown in Fig~\ref{fig:dtheta_theta}a) to obtain the fit coefficient $A  = (4.8\pm0.3)\times10^{-4}$ rad/s. By equating this with the first term of Eq.~\ref{eqn:dthetadt} we obtain $a= A\times(12R_e^{-1/2} + 1)/(2\Omega\cos\phi) \simeq 5.0$.  We see that $a$ is somewhat larger than unity, but not excessively so given the many approximations made in the model. It is unclear which approximations would lead to such an underestimation of the preferred orientation forcing.   The preferred orientation $\theta_m$ is given by $\sin\theta_m = B/A = 0.98\pi$ and is consistent with those obtained from the peak of $p(\theta_0)$ shown in Fig.~\ref{fig:theta_pref} for small $R$ in the large sample. This confirms that the $\theta_0$-dependent forcing is responsible for the non-uniform $p(\theta_0)$ in this parameter range.  This agreement is significant because there was no free parameter in Eq.~\ref{eqn:dthetadt} for determining $\theta_m$ ($A$ is determined mainly by the sinusoid amplitude and has only a small effect on $\theta_m$).   The fit amplitude $A$ vs. $R$ is shown in Fig.~\ref{fig:dtheta_theta}b, along with the model value for $A$ from Eq.~\ref{eqn:ab} with $a=5.0$, and with $R_e = 0.0345R^{1/2}$  which was found in other independent measurements \cite{BFA06} for the range of $R$ studied here.  The  large scatter around the fit can be qualitatively attributed to asymmetries of the sample (see Sect.~\ref{sec:asymmetry}). The scatter also prevents us from accurately testing the $R$-dependence of the model at this point.

We note that the line in Fig.~\ref{fig:dtheta_theta}b is nearly horizontal, indicating a very small $R$ dependence of the model.  This is because the denominator of $A$ in Eq.~\ref{eqn:ab} is dominated by the constant term equal to $1$; the term $12R_e^{-1/2}$ ranges from $0.1$ to $0.4$ in the relevant range of $R$. Physically this implies that drag is relatively unimportant in slowing down the azimuthal rotation of the LSC, especially at higher $R$. Thus the rotation rate is limited mainly by the reduction of the Coriolis force in proportion to $-\dot \theta$, {\it i.e.} by the rotation of the LSC itself in the retrograde direction.

\subsection{The probability distribution of the orientation}
\label{sec:ptheta}

Calculating the net rotation rate $\Omega_{\cal R}$, the total number of revolutions per unit time $\Omega_N$, and the probability distribution $p(\theta_0)$ from the model is more complicated because these properties are controlled by the turbulent ``noise" that causes the LSC orientation to meander.  In Sect.~\ref{sec:diff} we reported that the orientation meanderings have the diffusive quality that the root-mean-square rotation rate $\dot\theta_n^{rms}$ is given by $\dot\theta_n^{rms} = \sqrt{D_{\theta}/(n\delta t)}$, where $D_{\theta}$ is a diffusion constant. We thus adopt the idea that $\theta_0$ carries out a diffusive motion,  driven by the background turbulence, in the  potential created by the Coriolis force.  From Eq.~\ref{eqn:dthetadt} we find this potential to be given by

\begin{equation}
V \equiv -\int\dot\theta_0(\theta_0) d\theta_0 = A\cos\theta_0 +B\theta_0 \ .
\label{eqn:potential}
\end{equation}

\noindent With no noise, the orientation would become locked in a local minimum of this washboard potential and there would be no net rotation.    With strong noise, the local maxima in the potential become insignificant and the net rotation rate is equal to $-B$.  

To calculate both $p(\theta_0)$ and the net rotation rate for an intermediate noise strength we follow the method of Ambegaokar and Halperin  who solved a mathematically equivalent problem for the voltage across a Josephson junctions driven by a current source \cite{AH69}.  We start with a Fokker-Planck (or Smoluchowski) equation

\begin{equation}
 \frac{dp(\theta_0)}{dt} = \nabla[-p(\theta_0)\dot\theta_0(\theta_0) + D_{\theta}\nabla p(\theta_0)] \ .
\label{eqn:fokkerplanck}
\end{equation}

\noindent The probability distribution evolves in time according to the sum of an advective and a diffusive probability current.  This equation is valid in the case of strong damping, specifically where the diffusion velocity  $3UD/4L \stackrel{>}{_\sim} 13\Omega$ is large compared to the mean drift velocity (less than $\Omega\sin\phi$), and the mean free path $4LD/3U \approx 0.2$ rad is small compared to the distance between peaks of the potential ($2\pi$ rad).  In the stationary state the left side of Eq.~\ref{eqn:fokkerplanck} is equal to zero, and by integrating the right side we obtain

\begin{equation}
-p(\theta_0)\dot\theta_0(\theta_0) + D_{\theta}\nabla p(\theta_0) = -\frac{\omega_{\cal R}}{2\pi}\ .
\label{eqn:fokkerplanck2}
\end{equation}

\noindent Here $-\omega_{\cal R}/(2\pi)$ is a constant of integration whose value can be confirmed by integrating Eq.~\ref{eqn:fokkerplanck2} once more over the range from $0$ to $2\pi$ and using the fact that $p(\theta_0)$ is normalized with periodic boundary conditions in this reduced range.  In other words the net rotation rate  is the balance of the advective and diffusive terms.

Solving Eq.~\ref{eqn:fokkerplanck2} for $p(\theta)$ is done by assuming 

\begin{equation}
p(\theta_0) = \frac{\omega_{\cal R}}{2\pi D_{\theta}}f(\theta_0)g(\theta_0) 
\label{eqn:pthetadef}
\end{equation}

where 

\begin{equation}
f(\theta_0) \equiv \exp(-V/D_{\theta}) \ .
\label{eqn:ftheta}
\end{equation}

\noindent Here $p(\theta_0) \propto f(\theta_0)$ happens to be the solution for $B=0$, and we note that Eq.~\ref{eqn:potential} implies $-\nabla V = \dot\theta_0$. We apply periodic boundary conditions and get 

\begin{equation}
g(\theta_0) = \frac{1}{f(2\pi)-f(0)}
\left[f(0)\int_0^{\theta_0}\frac{d\theta'}{f(\theta')} + f(2\pi)\int_{\theta_0}^{2\pi}\frac{d\theta'}{f(\theta')}\right]
\label{eqn:ptheta}
\end{equation}

\noindent We next integrate Eq.~\ref{eqn:pthetadef} from 0 to $2\pi$. Requiring $p(\theta_0)$ to be normalized, we then have \cite{AH}

\begin{equation}
\omega_{\cal R} = 2\pi D_{\theta}
\left[ \int_0^{2\pi}d\theta_0 f(\theta_0)g(\theta_0)\right]^{-1}
\label{eqn:omegar_fp}
\end{equation}

\begin{figure}
\includegraphics[width=3in]{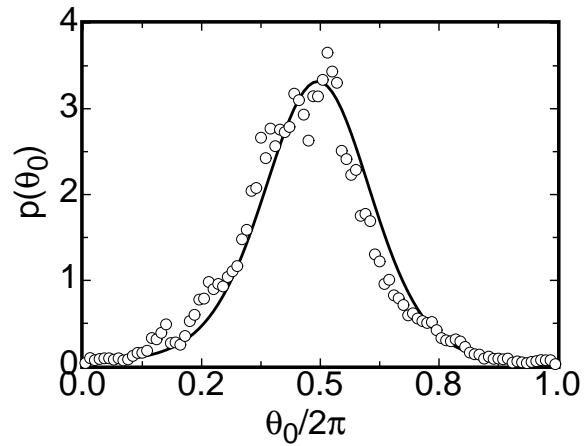}
\caption{The probability distribution of the orientation $p(\theta_0)$ for $R=5\times10^9$ in the large sample.  Open circles:  data.  Solid line: prediction based on the Fokker-Planck equation and the experimentally determined diffusivity $D_\theta$ without any free parameters. }
\label{fig:ptheta_fp}
\end{figure}

\noindent These equations must be solved numerically to obtain $\omega_{\cal R}$ and $p(\theta_0)$.  Figure \ref{fig:ptheta_fp} shows $p(\theta_0)$ for $R=5\times10^9$ in the large sample along with the result calculated from Eq.~\ref{eqn:pthetadef} using the experimentally determined values of $D_{\theta}$ given in Fig.~\ref{fig:D}  in Sect.~\ref{sec:diff}. There is good agreement without requiring any adjustable parameters.   Sample asymmetries at larger $R$ required the use of experimentally measured values for $A(R)$ in place of Eq.~\ref{eqn:ab} and a shift in $\theta_0$. With these adjustments, similarly good agreement was found at all $R$ studied in both samples.

\subsection{Net rotation rate}
\label{sec:rotation}

\begin{figure}
\includegraphics[width=3in]{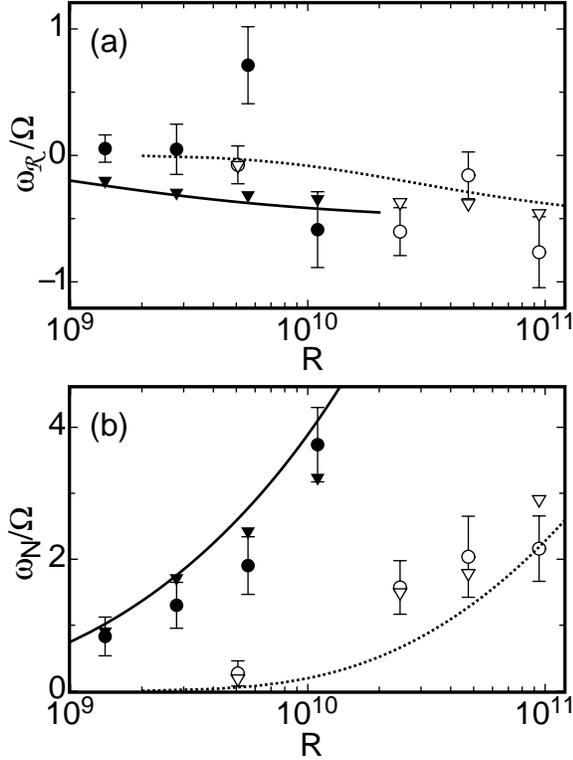}
\caption{(a):The net rotation rate $\omega_{\cal R}$ of the LSC vs. $R$.  Solid circles:  medium sample data.  Open circles:  large sample data.  Solid line:  model calculation of $\omega_{\cal R}$ with $A$ given by Eq.~\ref{eqn:ab} and $a = 5.0$ for the medium sample.  Dashed line:  model value of $\omega_{\cal R}$ for the large sample.  Solid triangles:  calculation of $\omega_{\cal R}$ for the medium sample where the potential barrier $A(R)$ is obtained experimentally  at each $R$ as discussed in Sect.~\ref{sec:inst_rot_rate}. Open triangles:  calculation of $\omega_{\cal R}$ for large sample with experimentally obtained $A(R)$. Because net rotations are infrequent, the data for $\omega_{\cal R}$ (circles) were averaged over multiple runs in a small range of $R$.  (b):  The rate of revolutions $\omega_{N}$ vs. $R$. Symbols have the same meaning as in part a.  The calcuations of $\omega_{N}$ are from Eq.~\ref{eqn:omegan}.   }
\label{fig:revolutions}
\end{figure}

Equation \ref{eqn:omegar_fp} can be used to numerically calculate the net rotation rate $\omega_{\cal R}$.  
The experimental data for $\omega_{\cal R}$ are shown in Fig.~\ref{fig:revolutions}a (circles).  The predictions from the model with $a=5.0$ in Eq.~\ref{eqn:ab} and $D_{\theta}$ from Eq.~\ref{eqn:D_theta} are shown as well (lines).  The agreement is less than perfect. Since the Coriolis force was apparently not the only asymmetry contributing to the potential barrier $V$, we also calculated $\omega_{\cal R}$ by substituting the measured values of $A(R)$ and $D_{\theta}(R)$for Eq.~\ref{eqn:ab}  and obtained the triangles in Fig.~\ref{fig:revolutions}a. The agreement with the measurements (circles) is improved somewhat.  In either case, the model predicts $\omega_{\cal R}$ close to $-0.3\Omega$ with some variation with $R$ for both samples.  While there is overall consistency with the experimental data from the large sample, reproducing correctly both the direction and magnitude of the net rotation, the model does not explain why we see no statistically significant net rotation in the medium sample.

It is interesting to note that the prediction of $|\omega_{\cal R}|$ increases with $R$.  This does suggest that there is a lower cutoff for $R$ below which net rotation cannot be observed,  with the cutoff depending on the resolution of the experiment.  For example at $R=5\times10^9$ in the large sample, the prediction is $\omega_{\cal R} = -0.03$, which is too small for us to distinguish from zero.   However, in our experiments cells of different heights were used to cover different ranges of $R$, and because of the factor of $L^2$ in $D$, increasing $R$ by increasing $L$ (and keeping $\Delta T$ constant) leads to an increase in $\omega_{\cal R}$ with decreasing $R$ in our parameter range.  Thus even though the theory predicts a cutoff $R$ below which net rotation cannot practically be observed, it does not explain the absence of net rotation in our  medium sample.  The limit of the net rotation rate as $D_{\theta}/A$ becomes large at large $R$ in both samples is  $\omega_{\cal R} = -B \simeq -0.5\Omega$.

\subsection{Frequency of revolutions}

Net rotations are very rare and thus the  experimental uncertainty of their rate of occurrence is relatively large. A stricter test of the model is achieved by calculating the total number of revolutions $\omega_N$.  To calculate $\omega_N$ from the model we follow the method of Kramers \cite{Kr40}.  
He calculated the rate of crossing a potential barrier for large barriers.  While in our case the diffusion constant $D_{\theta}$ is of the same order as and sometimes larger than $A$, we can still calculate the ratio $\omega_N/\omega_{\cal R}$ for these intermediately sized potential barriers.

In our model $\theta_0$ diffuses in a washboard potential with a series of minima. We would like to make use of the Fokker-Planck equation for the non-equilibrium probability distribution $p_m(\theta_0)$ of the LSC trapped in the $m^{th}$ potential well.   While our system is not in the large-potential-barrier limit of the Arrhenius-Kramers problem, the barriers are large enough to keep the LSC in a single well on average for several hours ($2\pi/\omega_N$) as can be seen, for example, in Fig.~\ref{fig:theta_t}. We claim that the barriers are large enough so that $p_m(\theta_0)$ is {\em nearly} stationary and the stationary Fokker-Planck equation (Eq.~\ref{eqn:fokkerplanck2}) is a good approximation in this case.  Since $p_m(\theta_0)$ is not periodic, we can treat two potential barriers $\Delta V_+$ and $\Delta V_-$ (which have different sizes because of the net rotation term $B$ in the potential) separately to obtain the rates $\omega_{\pm}$ of transition from one potential minimum $m$ to an adjacent minimum $m\pm$.   We rearrange Eq.~\ref{eqn:fokkerplanck2} for a stationary distribution, replacing $p(\theta_0)$ with $p_m(\theta_0)$ and $\omega_{\cal R}$ with $\omega_{\pm}$ to obtain

\begin{equation}
f(\theta_0)\omega_{\pm} \approx - 2\pi D_{\theta}\nabla[p_m(\theta_0)f(\theta_0)] 
\label{eqn:fokkerplanck_well}
\end{equation}

\noindent for $\theta_0$ between  $m$ and $m\pm$. Next we integrate over the potential barriers from $m$ to $m\pm$ to obtain

\begin{equation}
\omega_{\pm} = 2\pi D_{\theta}\frac{[p_m(\theta_0)f(\theta_0)]_{m\pm}^m}{ \int_m^{m\pm}f(\theta_0)d\theta_0} \ .
\label{eqn:rate_kramers}
\end{equation}
 
\noindent We note that $f(\theta_0)$ has the significance of the Boltzmann factor in the Kramers approach. We do not attempt to evaluate $p_m(\theta_0)$ for small potential barriers because we only need to evaluate the ratio $\omega_+/\omega_-$.  Since $p_m(\theta_0)$ represents a particle trapped near $m$, it is negligible near $m\pm$.  Now the numerator has the same magnitude in both the $+$ and $-$ cases and we can calculate the ratio of transition rates. Making use of Eqs.~\ref{eqn:ftheta} and \ref{eqn:potential} to evaluate the ratio of the remaining integrals we obtain 

\begin{eqnarray}
\frac{\omega_+}{\omega_-} &=& \frac{\int_{m-}^{m}f(\theta_0)d\theta_0}{\int_m^{m+}f(\theta_0)d\theta_0} \nonumber\\
&=& \frac{\int_{m-}^{m}f(\theta_0)d\theta_0}{\int_{m-}^{m}f(\theta_0+2\pi)d\theta_0} = \exp\frac{-2\pi B}{D_{\theta}} \ .
\label{eqn:rate_ratio}
\end{eqnarray}

\noindent With this the ratio $\omega_N/\omega_{\cal R}$ is found to be

\begin{equation}
\frac{\omega_{N}}{\omega_{\cal R}} \equiv \frac{\omega_+ + \omega_-}{\omega_+ - \omega_-} = -\coth\frac{\pi B}{D_{\theta}} 
\label{eqn:omegan}
\end{equation}

We show the data for $\omega_N$ in Fig.~\ref{fig:revolutions}b in a similar manner to the data for $\omega_{\cal R}$.  The model calculation for $\omega_N$ (lines) uses the value of $\omega_{\cal R}$ from Eq.~\ref{eqn:omegar_fp}, with $A$ obtained from Eq.~\ref{eqn:ab}.    The model captures the general trends of the data well.  When the experimentally obtained $A(R)$ are used in place of Eq.~\ref{eqn:ab} (triangles) there is excellent agreement between the measured values of $\omega_N$ and the model predictions in both samples.  

Since the potential tilting parameter $B$ increases the size of $\Delta V_+$ while equally decreasing the size of $\Delta V_-$, the value of $B$ has a negligible effect on $\omega_N$.   On the other hand the difference betwene $\Delta V_+$ and $\Delta V_-$ is responsible for the net rotation, so $B$ has a significant effect on $\omega_{\cal R}$.  Since the model predicts the correct $\omega_N$ for both samples, the disagreement between the prediction of $\omega_{\cal R}$ and the data in the medium sample suggests that the values of $B$ we used for the medium sample are suspect.  As seen from Eq.~\ref{eqn:ab}, the value of $B$ is essentially independent of experimental measurements. Thus the disagreement for the medium sample does not imply a shortcoming of the Fokker-Planck description but is an indication of unexpected physics. 

We note that the Kramers approach relies on the existence of (preferably large) potential barriers.  According to our model, the potential barriers disappear completely when $B>A$, which is the case for the Earth's Coriolis force near the north or south pole, or for intentionally  rotating experiments.  In these cases our calculation of $\omega_N$ would not apply.

\subsection{Comparison with other work}
\label{sec:otherwork}

Another experiment \cite{XZX06} similar to ours, at $\phi = 22^{\circ}$,  $\Gamma = 1$, $\sigma \approx 5$, $10^9 < R < 10^{10}$, and $L = 19.5$ cm,  found a clockwise net rotation at a rate that increased with $R$, which we estimate as an average over all $R$ represented in Fig. 4 of Ref.  \cite{XZX06} to be $\omega_{\cal R} \approx (-1.3 \pm 0.4)\Omega$.  Although the runs were much shorter and thus the uncertainties greater than ours, the net rotation rate reported was larger than any found by us, even though the measurements were in the range of $R$ and $L$ where we found no significant net rotation.  More alarming is that the magnitude of the rotation rate was larger than $\Omega\sin\phi =0.37\Omega$, the maximum allowed by the Coriolis force at this latitude.  Since the time series of the orientations showed a strongly locked preferred orientation, one would expect that the potential barriers would have significantly reduced $\omega_{\cal R}$ from the azimuthal drift rate, as was the case in our experiment. We conclude that our model is not consistent with the data of Ref. \cite{XZX06}, and it does not seem likely that any Coriolis-force-based model could account for the large rotation rate observed in that work.  

 An additional similar experiment from the same laboratory \cite{SXX05} at $R=5.3\times10^{10}$, $\sigma = 5.3$ but with aspect ratio $\Gamma = 0.5$ was seen to have a much faster average net rotation rate of $\omega_{\cal R} = +13.5\Omega$.  This flow was visualized to show that there was a single LSC roll, so our model should apply to that experiment with some minor modification.  In this case the absolute rate of rotation was very large, and the rotation was counter-clockwise when viewed from above whereas a Coriolis-force model would predict a much smaller rate in the clockwise direction for any experiment in the northern hemisphere. These facts suggest that it is not the Earth's Coriolis force that is responsible for the net rotation in that case.

We can compare the predictions of the model with the results of Hart et.~al.~\cite{HKO02}.  They deliberately rotated their samples in the laboratory frame. Since their axis of rotation was along the central vertical axis of the cell and their rotation rate  was much larger than one revolution per day, effectively $\phi = \pi/2$ and the Earth's Coriolis force may be neglected. Since the rotation rate was relatively fast, one would expect any potential barriers to be less relevant, and in any case those due to the Coriolis force vanish for $\phi = \pi/2$. Thus our calculation applicable to their system is that of the azimuthal drift rate with $\sin(\phi) = 1$ which gives $\omega_N = \omega_{\cal R} = -B = -\Omega/(12R_e^{-1/2}+1) \simeq -0.9\Omega$. The sign of this prediction agrees with but the magnitude is larger than that of the experimental result $\omega_{\cal R} = -0.23\Omega$.  Hart et.~al.~ did not see any net rotation below $R = 9\times10^9$, which they achieved with a smaller sample about the same size as our medium sample. This is in striking similarity to our own observations, and suggests there is a physical reason related either to the size of the sample or to $R$ that no net rotation was observed in these experiments.

Hart et al. \cite{HKO02} also developed a model to explain their results that is similar to ours in that the Coriolis force was balanced by drag. In the limit of zero drag, called the ``inertial oscillation" model by Hart et al., both models yield $\omega_{\cal R} = -\Omega$. Better agreement with the experiment can be achieved for a dissipation level significantly larger than that provided by the viscous drag across a laminar boundary layer assumed by us.  Indeed Hart et al. invoked an effective turbulent viscosity and introduced a greater level of dissipation into their model, thereby achieving good agreement with their measurements of $\omega_{\cal R}$.

\section{Effect of azimuthal sample asymmetry on the preferred orientation}
\label{sec:asymmetry}

At large $R$  we found a preferred orientation that was locked in the sample reference frame (see Fig.~\ref{fig:theta_pref}) and thus must have been due to some asymmetry of the apparatus.    We identified the top-plate cooling-system as the most likely source. This system consisted of a channel in the form of a double spiral with inlet and outlet at opposite ends of a diameter. \cite{BNFA05} As coolant traversed through the channel, its temperature increased slightly as it absorbed heat from the plate.  This meant that the coolant was coolest near the inlet and warmest near the outlet.  We confirmed this by conducting separate experiments with coolant flowing in opposite directions. In both cases the temperature monitored at five locations in the top plate near the fluid-solid interface showed that here was a small lateral temperature gradient in the top plate that was positive from the inlet to the outlet. In both cases  the preferred orientation $\theta_m$ at large $R$ corresponded to the warmer side of the plate. 

The double-spiral design had been intended to provide a uniform temperature over the entire top plate because ideally the average temperature of adjacent channels should be the same everywhere. However, the experimental results suggest that this cancellation between adjacent channels was not sufficient to prevent a biased azimuthal orientation of the very sensitive LSC. We believe that the incomplete cancellation near the inlet and outlet may be responsible for the problem. The cooling of the top plate  contrasts with the heating of the bottom plate, where the same amount of heat is dissipated per unit length along the entire length of the heater wire and where the spacing between adjacent lengths of the wire was smaller than that between the spiral channels.  

In the remainder of this section  we model the effect on the LSC orientation of a small horizontal temperature gradient in the top plate due to the cooling-system geometry. 
First we focus on the horizontal temperature different $\delta T$ in the top plate between two thermistors a distance $3L/4$ apart along a diameter nearly passing through the inlet and outlet.  The magnitude in the large sample for $R \stackrel{>}{_\sim} 2\times10^{10}$ was given approximately by $\delta T/\Delta T = 1.2\times10^{-8} R^{0.57}$.  A top-plate thermistor temperature and the side-wall thermistor temperature below it showed a positive correlation, with the top-plate temperature leading the side-wall temperature, suggesting that the top-plate temperature determined the LSC orientation and not the other way around.   For smaller $R$ the temperature gradient fell below the fit and even became slightly negative from the inlet to the outlet.  At these $R$ the Coriolis force was shown to determine $\theta_m$, and the top plate happened to be aligned so that the cooling system pushes the LSC in the opposite direction as the Coriolis force,  so the temperature measurements suggest that the LSC determined the top-plate temperature-profile at small $R$.  

Next, to model the effect of the top-plate asymmetry on the LSC, we will assume that the top-plate temperature varies linearly along a diameter so that its magnitude is given by $4r\delta T/(3L)|\cos(\theta-\theta_0)|$.  If this $\delta T$ is added to the thermal boundary-layer temperatures, then it adds an additional buoyant force to the boundary layer of which a component proportional to $\sin(\theta_c-\theta_0)$ will orient the LSC towards $\theta_c$ with the forcing
$$\dot u_{\theta} \approx g\alpha \frac{4r\delta T}{3L} |\cos(\theta-\theta_0)|\sin(\theta_c - \theta_0)\ .$$

\noindent Now we take an azimuthal average of the acceleration over the entire sample by doing the integral of $\dot u_{\theta}$ over $\theta$ to get a factor of $2/\pi$. Multiplying by the fractional thermal boundary-layer width $l/L \approx 1/(2{\cal N})$ one obtains the angular acceleration 

\begin{equation}
\ddot\theta_0 = \frac{\left<\dot u_{\theta}\right>}{\left<r\right>} \approx \frac{4R\nu^2}{3\pi\sigma {\cal N} L^4}\frac{\delta T}{\Delta T}\sin(\theta_c - \theta_0) \ .
\label{eqn:cooling}
\end{equation}

 \begin{figure}                                                
\includegraphics[width=3in]{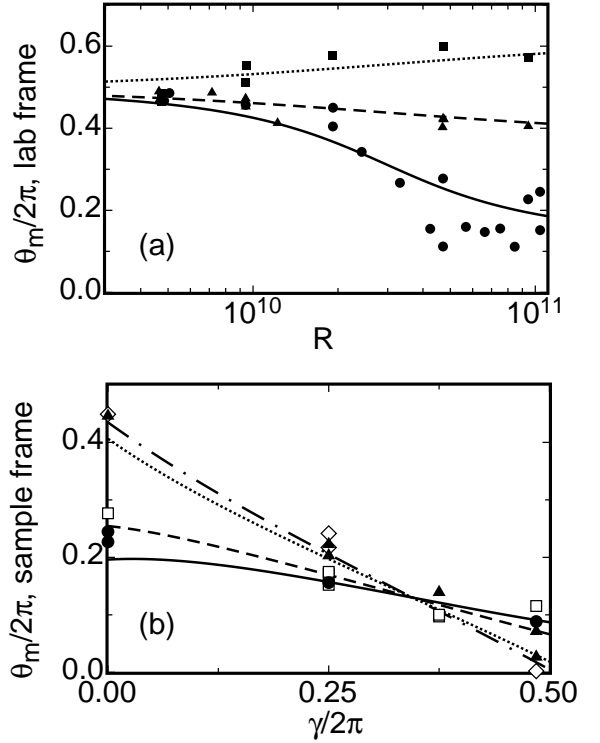}
 \caption{(a):  The preferred orientation $\theta_m$ vs. $R$ in the laboratory frame for various rotation angles $\gamma$ of the large apparatus. Solid circles: $\gamma = 0$, fit by solid line.  Solid triangles:  $\gamma = 0.50\pi$, fit by dashed line.  Solid squares:  $\gamma = 0.97\pi$, fit by dotted line.  The fitting function represents $\theta_m$ due to a combination of the Coriolis force and the top plate asymmetry.  (b):  Much of the same data plotted as $\theta_m$ vs. $\gamma$ for several values of $R$ in the frame of the large sample.  Points represent data, while the lines represent the model calculation using the fit values from Fig.~a.  Solid circles and solid line:  $R = 1\times 10^{11}$.  Open squares and dashed line:  $R = 4.7\times 10^{10}$.  Solid triangles and dotted line:  $R=9.4\times 10^9$. Open diamonds and dashed-dotted line:  $R=5\times 10^9$.}
 \label{fig:theta_pref_cool}                                       
\end{figure}

\noindent Using the results ${\cal N} = 0.0602R^{1/3}$ \cite{BNFA05} and $R_e = 0.0345R^{1/2}$ \cite{BFA06} of prior measurements for our samples, we see that the acceleration scales roughly as $R^{1.24}$, compared to the Coriolis-force acceleration from Eq.~\ref{eqn:coriolis} which scales as $R^{0.5}$; so this explains why the Coriolis force determines $\theta_m$ at small $R$ but the cooling system determines $\theta_m$ at large $R$.  To find the new preferred orientation, we set the forcings given by Eq.~\ref{eqn:coriolis} and Eq.~\ref{eqn:cooling} equal to each other, setting $\dot\theta_0 = 0$, and inserting $\gamma$ in the Coriolis-force term to account for the different apparatus orientations. We obtain a relationship between $R$ and $\theta_m$ given by
\begin{equation}
\left(\frac{R}{R_t}\right)^{0.74} = \frac{\sin(\theta_m-\gamma)-({1}/{2a})\tan\phi}{\sin(\theta_m-\theta_c)} \ .
 \label{eqn:rcrit}
 \end{equation}
 Here $R_t$ is the transitional Rayleigh number where the cooling-system asymmetry overtakes the Coriolis force in determining $\theta_m$. From the numbers given we calculate $R_t \simeq 3.6\times 10^{10}$.  We fitted our data for $\theta_m$ vs. $R$ from the large sample for several apparatus orientations $\gamma$ to Eq.~{\ref{eqn:rcrit}. This fit is shown in Fig.~\ref{fig:theta_pref_cool}a and gave $R_t = 3\times10^{10}$ and $\theta_c = 0.26\pi$.    With these parameters, we also calculate $\theta_m$ vs. $\gamma$ from the model for several values of $R$. The results are  shown in Fig.~\ref{fig:theta_pref_cool}b for a different perspective on much of the same data as those shown in Fig.~\ref{fig:theta_pref_cool}a.  The transitional Rayleigh number $R_t$ is predicted quite accurately for an order-of-magnitude model.  The large-$R$ limit of the preferred orientation $\theta_c$ is opposite the coldest top-plate thermistor, which is the first thermistor the fluid passes in the cooling channel, so these data are consistent with the cooling system being responsible for determining $\theta_m$ at large $R$ in our experiments.   The cooling system orientation $\theta_c$ was shown in Fig.~\ref{fig:theta_pref_fit}b.  In that figure, the background of $p(\theta_0)$ for the $\gamma=0$ data is notably larger than for the other $p(\theta_0)$ shown, which can be explained by the fact that at $\gamma = 0$, the Coriolis force and the top-plate asymmetry are trying to force the LSC into two opposing preferred orientations and the partial cancellation of these forces results in a more uniform $p(\theta_0)$.  The variation of the potential barrier $A$ with $R$ as shown in Fig.~\ref{fig:dtheta_theta}b can also be explained qualitatively.  Since the cooling-system forcing and the Coriolis force oppose each other, the net effect is to reduce the net potential barrier at large $R$ in the large sample from the Coriolis force prediction.  A horizontal temperature gradient across the top plate from the medium sample was also measured at large $R$ to go from the coolant outlet to the inlet.  In this case the top plate was placed with the inlet on the east side, so that the cooling system forced the LSC in the same direction as the Coriolis force.  This can be seen to increase the potential barrier $A$ relative to the Coriolis force prediction for the medium sample in Fig.~\ref{fig:dtheta_theta}b.

\section{Cancellation of preferred orientation by tilt}
\label{sec:tilt}

\begin{figure}
\includegraphics[width=3in]{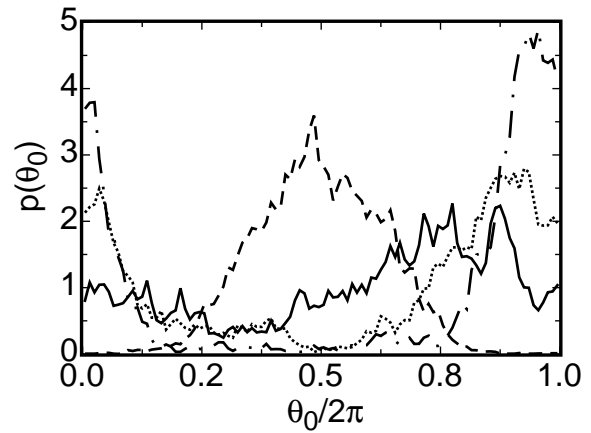}
\caption{ The probability distribution of the orientation $p(\theta_0)$ for several tilt angles at $R=9.0\times10^{9}$ in the large sample.  Dashed line: $\beta = 0$.  Solid line: $\beta = 0.0035$ rad.  Dotted line:  $\beta = 0.0070$ rad.  Dashed-dotted line: $\beta = 0.0105$ rad.  Notice that $p(\theta_0)$ is most nearly uniform for $\beta = 0.0035$ rad.}
\label{fig:prob_theta_tilt}
\end{figure}

We can tilt the apparatus to an angle $\beta$ relative to the gravitational acceleration, defined to be positive when the usually cooler side of the top plate is raised so as to add buoyancy to the boundary layers to oppose the prevailing preferred orientation.  At an appropriate angle $\beta_0 > 0$, we expect the effects of the various asymmetries to be cancelled out by the added buoyancy due to the tilt.  It would be a simple matter to combine the models presented above for the Coriolis force and cooling-system force with the tilt model presented in Ref. \cite{ABN05} to estimate $\beta_0$, but in the interest of brevity we show only the experimental results.  The probability distribution $p(\theta_0)$ is shown for several small positive tilt angles in Fig.~\ref{fig:prob_theta_tilt} at $R=9.0\times10^{9}$ in the large sample.    These experiments were done at $\gamma=0$, where the Coriolis force and cooling-system asymmetries oppose each other, and the Coriolis force dominates at the smaller $R$ while the cooling system dominates at the larger $R$.  For larger tilt angles $p(\theta_0)$ is sharply peaked near the southeast ($\theta_0/2\pi = 0.875$), which corresponds to the high side of the tilted sample, as we would expect.  At the intermediate tilt angle of $\beta = 0.0035$, we find the most uniform $p(\theta_0)$, suggesting a close balance between the initial asymmetries of the sample and the buoyancy effects due to tilting.  To get a precise value of $\beta_0$, we fit Eq.~\ref{eqn:ptheta_gauss} to $p(\theta_0)$ to obtain the width $\sigma_{\theta}$ and background constant $C$.   Each fitparameter $x$ was then fit to an empirical function $x(\beta) = x_0 +x_1\exp(-|\beta-\beta_0|/x_2)$ for $|\beta| < 0.021$ rad.  The values of $\beta_0$ from the fit are shown in Table 1.

\begin{table}
\begin{center}
\begin{tabular}{c|c|c}
parameter				&$R=9.0\times 10^{9}$ 			&$R=8.9\times 10^{10}$\\
\hline	
$\sigma_{\theta} [p(\theta_0)]$	&~~$0.0026\pm 0.0001$			&$0.0021 \pm 0.0001$\\
$C [p(\theta_0)]$			&~~$0.0031\pm 0.0002$			&$0.0014 \pm 0.0001$\\
$A$						&~~$0.0029\pm 0.0004$			&$0.0015 \pm 0.0004$\\
$\delta$					&$-0.0044 \pm 0.0009$			&$0.0056	\pm 0.0009$\\	
$\omega_{r}$				&$-0.0032 \pm 0.0011$			&$0.0072 \pm 0.0010$\\
$\dot\theta_{rms}$			&$-0.0031 \pm 0.0021$			&$0.0058 \pm 0.0039$\\
$R_e $					&$-0.0023 \pm 0.0010$			&$0.0058 \pm 0.0018$\\	
\end{tabular}
\label{tb:beta0}
\caption{The values of the tilt angle $\beta_0$ where the probability distribution of various measured parameters (left column) had a maximum for the large sample. Results are given at two values of $R$.  The preferred orientation at $\beta=0$ is determined by the Coriolis force at the smaller $R$, and by asymmetries of the top-plate cooling-system at the larger $R$. }
\end{center}
\end{table}

Several other  parameters were measured at various tilt angles to find $\beta_0$ at two values of $R$ in the large sample, with the results for $\beta_0$ shown in Table 1.   We previously reported the frequency of reorientations (spontaneous changes of the LSC orientation) $\omega_r$ vs. $\beta$, and showed that a Gaussian function fit to  $\omega_r(\beta)$ with a maximum at some $\beta_0$, which has the same physical meaning as before. We found $\beta_0 = 0.0093 \pm 0.0010$ rad at $R=9.4\times10^{10}$ in the large sample \cite{ABN05} and $\beta_0 = 0.0022 \pm 0.0006$ rad at $R=1.1\times10^{10}$ in the medium sample \cite{BA06}.    The other parameters measured were the coefficient $A$ representing the potential barrier strength; the LSC temperature amplitude $\delta$;  the root-mean-square rotation rate $\dot\theta_{rms}$;  and the Reynolds number $R_e$.  In each of these cases the parameter $x'$ is fit to an empirical function $x'(\beta) = x'_0 +x'_1|\beta-\beta_0|$ for $|\beta| < 0.021$ rad, with the value of $\beta_0$ again representing the balance between tilt and other asymmetric forces.  Since $\beta$ is defined positive when the tilt-induced buoyancy opposes the prevailing preferred orientation of the LSC, a positive value $\beta_0$ means that the prevailing asymmetries affect the measured parameter as expected.  At $R=8.9\times 10^{10}$, all of the values of $\beta_0$ are positive, so the cooling system asymmetry affects all of the measured parameters.  This is reasonable because the tilt model and the cooling-system model both assumed the mechanism of increased buoyancy in the boundary layer.  However, at $R=9.0\times 10^{9}$ some of the values of $\beta_0$ are negative, indicating that the prevailing asymmetry, in this case the Coriolis force, does not affect those measured parameters as strongly as the cooling system -- which also opposes the Coriolis force and thus enhanced the tilt effect at this $R$ -- even though the Coriolis force had a stronger effect in determining $\theta_m$ at this $R$.  The values of $\beta_0$ are smaller in magnitude for the parameters relating to $p(\theta_0)$ at both $R$ because the Coriolis force and cooling system asymmetries both apply and oppose each other, so there is a partial cancellation of forces.   

These data from the tilting experiments further confirm that the Coriolis force does affect $p(\theta_0)$ and the local azimuthal forcing $\langle\dot\theta_0(\theta_0)\rangle$ that is responsible for $p(\theta_0)$, as we had already shown by other methods, but they also show that the Coriolis force does not significantly affect $\delta$, $R_e$, $\omega_{r}$, or $\dot\theta_{rms}$.  Because the Coriolis force only deflects moving fluid and does not enhance or damp its speed, it is entirely reasonable that it does not directly affect $R_e$ and $\delta$.  The other values, $\omega_r$ and $\dot\theta_{rms}$, are presumably due to turbulence, but we do not know the detailed mechanisms.   We can not rule out the possibility of weaker, "higher order" couplings between the Coriolis force and the LSC.  At $R=8.9\times10^{10}$, the cooling system increases $R_e$ by an amount equivalent to  a tilt of $\beta = 0.006$ rad.  We previously reported that $R_e$ increases by 185\% per rad of tilt \cite{ABN05}, so our results suggest that the cooling system has the effect of increasing $R_e$ by about 1\% at this $R$. We measured ${\cal N}$ as well, but our resolution of 0.1\% is not enough to measure such small differences.  We have previously reported the sensitivity of ${\cal N}$ to tilt is a reduction of only 3.1\% per radian of tilt \cite{ABN05}, so even if we expected to find $\beta_0 = 0.006$ rad, that would imply a reduction in ${\cal N}$ of only 0.02\%.  However, we consider even this small effect from the Coriolis force unlikely, as much evidence exists that ${\cal N}$ is determined by the thermal boundary layers and not the LSC.

\section{Concluding remarks}
\label{sec:summary}

We identified several interesting effects of the Earth's Coriolis force and the turbulent background fluctuations on the large-scale circulation of turbulent Raleigh-B\'enard convection.  There are clockwise and counter-clockwise relatively sudden revolutions through $2\pi$ of the orientation of the LSC circulation plane. The clockwise ones are slightly more frequent than the counter-clockwise ones, leading to a net rotation with an average rate of less than one rotation per day. There is a preferred value and a distribution about it, sampled over time, of the LSC orientation. Both the preferred value  and the distribution can be understood quantitatively in terms of the diffusive LSC meandering in the Coriolis-force potential, driven by the turbulent fluctuations. This model agrees with experimental results extremely well for small Rayleigh numbers in one of our samples, but at larger $R$ a small asymmetry of the experimental apparatus has to be invoked to explain all the data. We gave a more complete summary of our findings in the Introduction and will not repeat this here. Instead, we make a few additional observations.

We used two samples: a ``medium" one over the Rayleigh-number range $3\times10^8 \stackrel{<}{_\sim} R \stackrel{<}{_\sim} 10^{10}$ and a ``large" one over the range $5\times10^{9} \stackrel{<}{_\sim} R \stackrel{<}{_\sim} 10^{11}$.  We saw no net rotation in our medium sample, and one might wonder why this was the case.  We suspect that the reason may be a change to a different flow structure with decreasing $R$ that responds differently to the Coriolis force.  A different flow structure is suggested based on images of the LSC reported by Ref.~\cite{XSZ03}.  Their experiments are done with $\Gamma=1$ and $\sigma=4$, with images taken by particle image velocimetry at $R=3.8\times10^9$ (which we can achieve in our medium sample) and $R=3.5\times 10^{10}$ (which we can achieve in our large sample).  The LSC at the higher $R$ is more square in shape, filling out the container, while the LSC at the smaller $R$ more nearly has the shape of an ellipse oriented with its long axis along a diagonal of a vertical sample cross section. In the latter case there are counter-rotating eddies at opposite  corners of the sample.  Since these smaller eddies rotate in the direction opposite the LSC, we would expect that the Coriolis force would have the opposite effect on them, which would partially cancel out the net effect on the LSC, and resulting in less net rotation than otherwise expected.  These eddies would also be expected to reduce $A$ and thus widen  $p(\theta_0)$ and increase $\omega_N$, but since our model calculations used the experimentally obtained $A(R)$ this does have any implications for the applicability of the Fokker-Planck description.

The calculation of $p(\theta_0)$ and the frequency of revolutions $\omega_N$ from the Fokker-Planck equation, the Coriolis-force potential, and the turbulence-driven diffusion  is remarkably accurate, which suggests that treating the LSC meanderings as a diffusive process in a potential is an excellent model for the system.  However, this does not completely describe the azimuthal dynamics of the LSC because it does not account for other spectacular events, in particular rotations and cessations.  We found previously by numerical simulation that a diffusive model underestimates the frequency of reorientations, even without potential barriers \cite{BA06}.  Cessations of the LSC have a duration of about one turnover time of the LSC, and rotations have a duration of less than 10 turnover times \cite{BA06}. The potential barriers identified are too weak to have a significant effect over such short time scales.  The fact that $D_{\theta}$, from time series with reorientations removed, leads to an accurate calculation of $\omega_N$ suggests the perhaps surprising result that rotations and cessations do not contribute significantly to $\omega_N$.  If we had used $D_{\theta}$ with reorientations included, then we would have over-estimated $\omega_N$ by about a factor of two.

Studies of the asymmetry of our top-plate cooling-system provided some useful information for future experiments.  While the LSC orientation and related dynamics have proven to be very sensitive to asymmetries of the system in general, we have shown that several aspects of the LSC, including $R_e$, $\delta$, $\dot\theta_{rms}$, and $\omega_r$, are not significantly affected by the Coriolis force, while they are affected by cooling-system asymmetries.  In many studies, the latter group of properties is a more physically interesting aspect of the LSC than $p(\theta_0)$, so while the Coriolis force may not affect the results of these studies, a poorly designed cooling system will, particularly at large $R$ where the temperature gradients in the cooling system become larger.  The problem we had was a cooling system in which the temperature drop of the coolant as it flowed through the cooling channel left an associated thermal imprint on the plate, which then affected the LSC temperature profile.   To minimize this effect,  the temperature drop in the coolant -- given by the ratio of heat dissipated by the coolant to the total heat capacity of the coolant -- should be small compared to the LSC temperature amplitude.  Alternatively, the cooling system should be designed with such a geometry that the temperature drop of the coolant does not cause a horizontal temperature gradient in the plate.

\section{Acknowledgments}

We are grateful to Werner Pesch for stimulating discussions. This work was supported by the US Department of Energy through Grant  DE-FG02-03ER46080.


\begin{references}
\bibitem{HKO}{The caption of Fig. 4 of \cite{HKO02} gives $R = 2.9\times 10^9$. We assume that this is a misprint since the values of the Taylor and Rossby numbers cited there, together with the Prandtl number, imply that $R \simeq 2.9\times 10^{11}$, and since the authors stated previously in their paper that there was no rotation of the LSC for $R$ as large as $9.2\times 10^9$.}
\bibitem{AH}{There appears to be a typo in Eq.~9 of Ref.~\cite{AH69}.  The limits of integration on the last integral over $\theta'$ should read from $\theta$ to $2\pi$ instead of $0$ to $2\pi$.}
\end{references}
\end{document}